\documentclass[twocolumn, a4paper, times, trackchanges]{aastex63}

\usepackage{linenofix}

\usepackage{amsmath}
\usepackage{graphicx}
\usepackage{natbib}

\newcommand{\lya}{Lyman-$\alpha$}
\newcommand{\iTot}{$I_{\text{tot}}$}

\begin{document}
\title{Update of the Solar Lyman-Alpha Profile Line Model}

\correspondingauthor{Izabela Kowalska-Leszczynska}
\email{ikowalska@cbk.waw.pl}

\author[0000-0002-6569-3800]{Izabela Kowalska-Leszczynska}
\affiliation{Space Research Centre PAS (CBK PAN),
Bartycka 18A, 00-716 Warsaw, Poland}

\author[0000-0003-3957-2359]{Maciej Bzowski}
\affiliation{Space Research Centre PAS (CBK PAN),
Bartycka 18A, 00-716 Warsaw, Poland}

\author[0000-0002-5204-9645]{Marzena A. Kubiak}
\affiliation{Space Research Centre PAS (CBK PAN),
Bartycka 18A, 00-716 Warsaw, Poland}

\author[0000-0002-4173-3601]{Justyna M. Sok{\'o}{\l}}
\affiliation{Space Research Centre PAS (CBK PAN),
Bartycka 18A, 00-716 Warsaw, Poland}
\affiliation{NAWA Bekker Fellow, Department of Astrophysical Sciences, Princeton University, Princeton, NJ 08544, USA}

\begin{abstract}
We present a modification of a model of solar cycle evolution of the solar \lya{} line profile, along with a sensitivity study of interstellar neutral H hydrogen to uncertainties in radiation pressure level. The line profile model, originally developed by \citet{IKL:18a}, is parametrized by the composite solar \lya{} flux, which recently was revised \citep{machol_etal:19a}. We present modified parameters of the previously-developed model of solar radiation pressure for neutral hydrogen and deuterium atoms in the heliosphere. The mathematical function used in the model, as well as the fitting procedure, remain unchanged. We show selected effects of the model modification on ISN H properties in the heliosphere and we discuss the sensitivity of these quantities to uncertainties in the calibration of the composite \lya{} series.
\end{abstract}

\section{Introduction}
\label{sec:intro}
\noindent
The solar resonant radiation pressure in the \lya{} spectral line is an important factor determining the distribution of the interstellar neutral H (ISN H) in the inner heliosphere \citep{tarnopolski_bzowski:09}. The density of ISN H near 1~au from the Sun (hence, consequently, the distribution of the heliospheric backscatter glow in the sky) and the flux of ISN H are sensitive functions of the magnitude of the solar radiation pressure. 

\citet{tarnopolski_bzowski:09} developed a model of the evolution of the solar \lya{} line profile integrated over the solar disk during the varying solar activity. This model was parametrized by the solar \lya{} composite flux, routinely measured by Laboratory for Atmospheric and Space Physics, University of Colorado (LASP), Boulder \citep{woods_etal:00}. The baseline data for the Tarnopolski \& Bzowski model were observations of the solar \lya{} line profile from SUMER/SOHO for a dozen dates during approximately half of the solar cycle \citep{lemaire_etal:02, lemaire_etal:05}. 

\begin{figure}
\centering
\includegraphics[width=0.95\columnwidth]{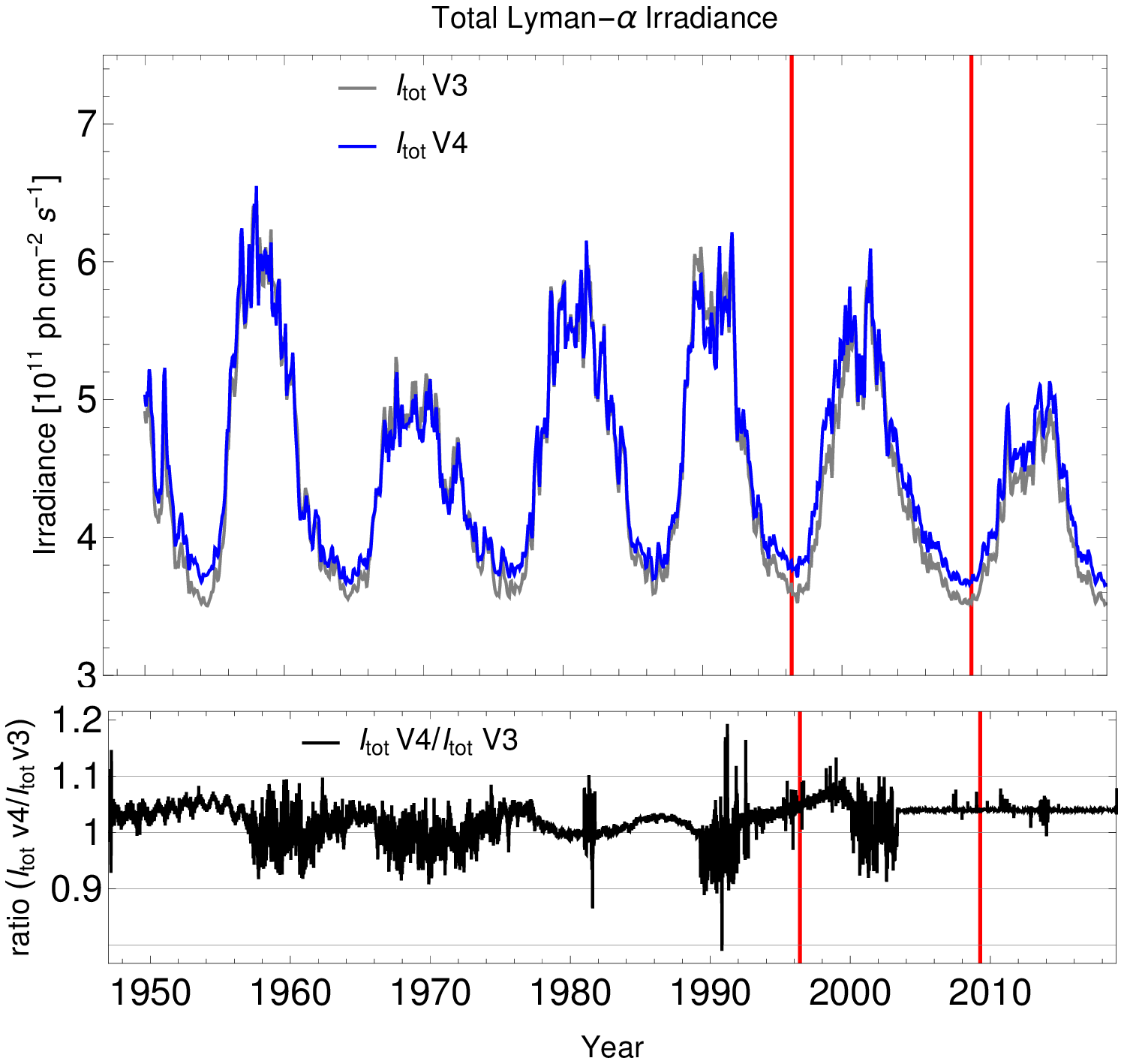}
\caption{Comparison of the two versions of the composite \lya{} flux from LASP: Version 3 (V3) and Version 4 (V4) (\citet{woods_etal:05a} and \citep{machol_etal:19a}, respectively). Top panel presents the data averaged over Carrington period for Version 3 in gray and for Version 4 in blue. Bottom panel shows the ratio of \iTot{} V4 to \iTot{} V3. Note that the change between Version 3 and Version 4 after $\sim 2005$ is by an almost constant factor of 1.04.
\label{fig:Itotv3v4}
}
\end{figure}

With observations of the line profile from more than forty dates covering a full solar cycle (1996--2009), published by \citet{lemaire_etal:15a}, \citet{IKL:18a} developed a more refined model of the dependence of this profile on the magnitude of the solar \lya{} composite flux, which will be referred to as the IKL model of radiation pressure. In this model, the line profile is composed of three main components: (1) a kappa-like general profile, (2) a Gaussian central reversal, responsible for the characteristic self-reversed structure with two horns, and (3) a linear background (foot). The parameters of the functions defining these components were assumed to be linear functions of the line- and disk-integrated \lya{} intensity, available as the LASP composite flux\footnote{Available from http://lasp.colorado.edu/lisird/data/composite\_lyman\_alpha/}. The original profiles observed by \citet{lemaire_etal:15a} and the model by \citet{IKL:18a} both used the same version of the composite \lya{} flux, namely Version 3 \citep{woods_etal:05a}.

\citet{machol_etal:19a} re-calibrated the composite \lya{} flux, using observations corrected based on an improved model of instrument aging and a more advanced method of filling the inevitable gaps in daily observations using different proxies. The resulting Version 4 of the \lya{} composite flux is compared with Version 3 in Figure~\ref{fig:Itotv3v4} as well as in Figure 5 in \citet{machol_etal:19a}. Version 4 of the composite flux has been improved by using as the reference data from SORCE SOLSTICE instead of those from UARS SOLSTICE, but also by using the solar radio flux F30 \citep{tanaka_kakinuma:57a} instead of the F10.7 \citep{tapping:13a} flux wherever possible. Also there was an issue with a 1~au correction of the F10.7 radio flux in Version 3 that is now removed. As a result, Version 4 is a major improvement with respect to Version 3. Typical differences between the two versions are $\sim \pm 10$\%. Generally, the magnitudes of the flux during the minimum of solar activity are somewhat higher in Version 4 than in Version 3. The ratio of the irradiances V4/V3 after 2005 is approximately constant and equal to 1.04, but for earlier dates it oscillates inside $\sim \pm 5$\%, with occasional departures to $\pm 10$\%, and sometimes even to $\pm 20$\% (several days in 1991). During the solar maxima, the ratio of the irradiances V4/V3 is rapidly changing, and the magnitude of the V4 flux is as often smaller as it is higher than in V3. 

\citet{IKL:18b} demonstrated a high sensitivity of the density and flux of ISN H near 1~au to details of the solar \lya{} line profile, studying differences between predictions of the Warsaw Test Particle Model \citep[nWTPM; ][]{tarnopolski_bzowski:09} run with radiation pressure models from \citet{tarnopolski_bzowski:09} or, alternatively, from \citet{IKL:18a}. The high sensitivity of the ISN H to details of radiation pressure inferred from this analysis stimulated us to update the model of \citet{IKL:18a} based on the updated time series of the LASP \lya{} composite flux and to investigate how this update modifies the ISN H inside the heliosphere. An added benefit from this analysis is an illustration of the sensitivity of the density of interstellar neutral hydrogen (ISN H) within a few au from the Sun is to the solar total irradiance, and of the non-linear nature of this sensitivity.

In the following, we re-evaluate the parameters of the radiation pressure model proposed by \citet{IKL:18a} for hydrogen and deuterium with the baseline solar \lya{} profiles updated by the new Version 4 \lya{} composite flux series. In Section \ref{sec:model} we start with re-normalization of the input line profiles from \citet{lemaire_etal:15a}. We then derive new coefficients of the model from \citet{IKL:18a}. In Section \ref{sec:comparison} -- following the approach by \citet{IKL:18b} -- we investigate the effect of the update of the model on selected aspects of ISN H inside the heliosphere. Finally, in Section \ref{sec:summary} we summarize and conclude our work.  

\section{Updated model of solar radiation pressure}
\label{sec:model}

\subsection{Renormalization of the profile line observations}
\label{sec:model:observations}

\noindent
In the original paper by \citet{lemaire_etal:15a}, the absolute scaling of the observed profiles was done by satisfying the requirement for the integrated spectral irradiance measured by SUMER to be equal to the magnitude of the composite \lya{} irradiance for the day of observation. The observed profiles were normalized using the total \lya{} irradiance Version 3. We re-scaled the \lya{} profiles published by \citet{lemaire_etal:15a} using the updated, Version 4, of the composite \lya{} flux. The rescaling factors are ratios of \iTot{} from Table A.1 in \citet{lemaire_etal:15a} to \iTot{} in Version 4:
\begin{eqnarray}
f_{rs}(t_i)=\frac{I_{\text{tot,V4}}(t_i)}{I_{\text{tot,V3}}(t_i)},
\end{eqnarray}
where $f_{rs}$ is the scaling factor for a day $t_i$, $I_{\text{tot,V3}}(t_i)$ and $I_{\text{tot,V4}}(t_i)$ are the total solar irradiance in Version 3 and 4, respectively, for $t_i$. We next multiplied the original line profiles by the appropriate $f_{rs}$ coefficients. 

\begin{figure*}
\centering

\includegraphics[width=0.46\textwidth]{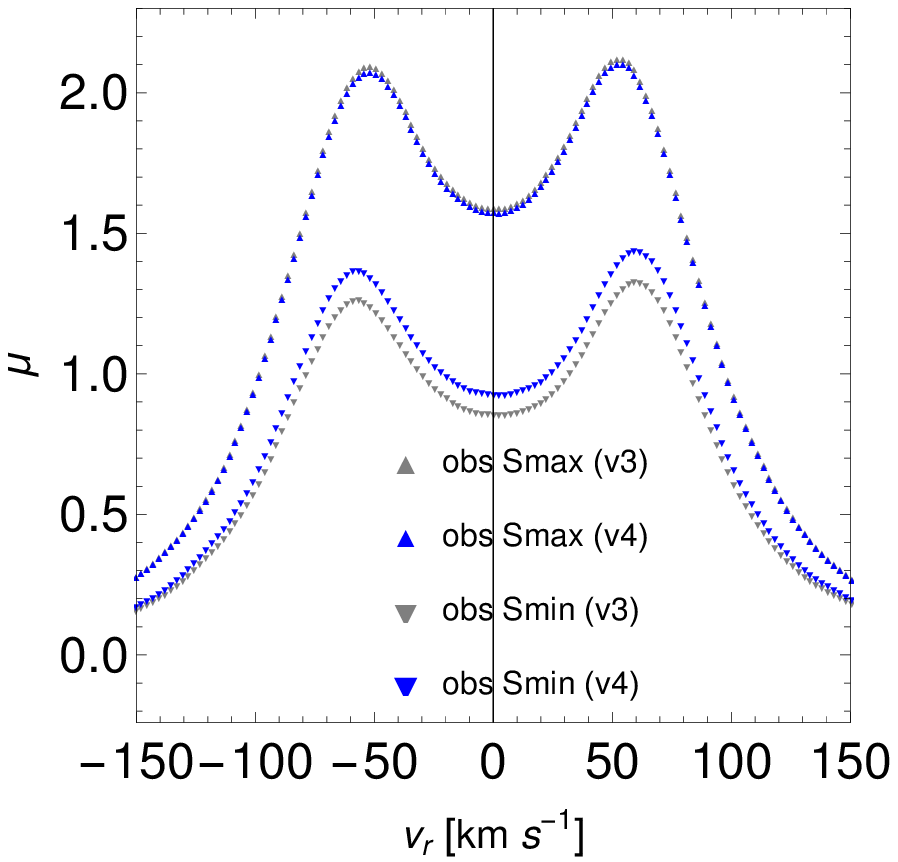}
\includegraphics[width=0.46\textwidth]{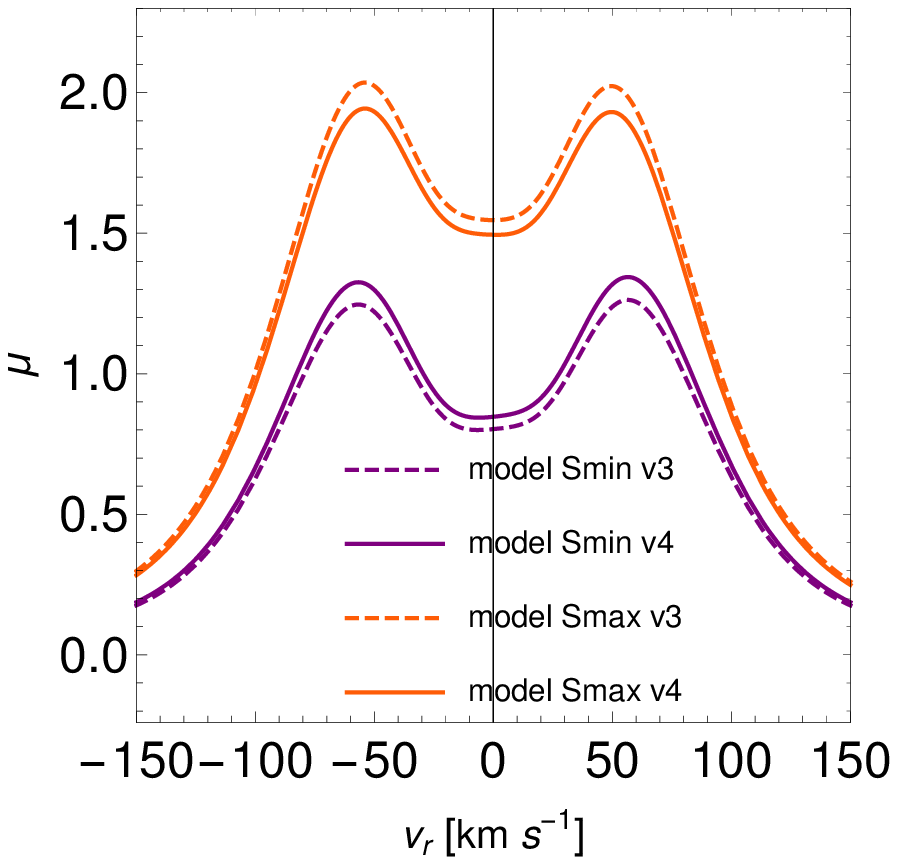}

\includegraphics[width=0.46\textwidth]{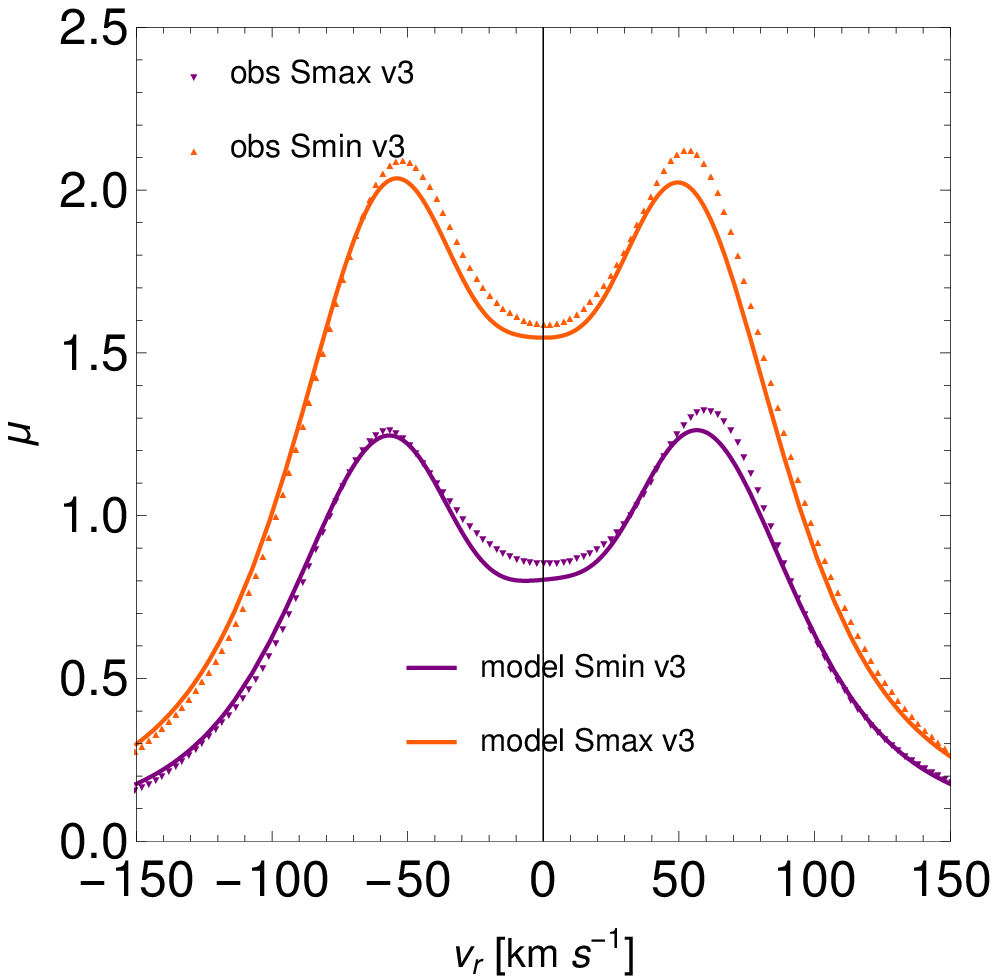}
\includegraphics[width=0.46\textwidth]{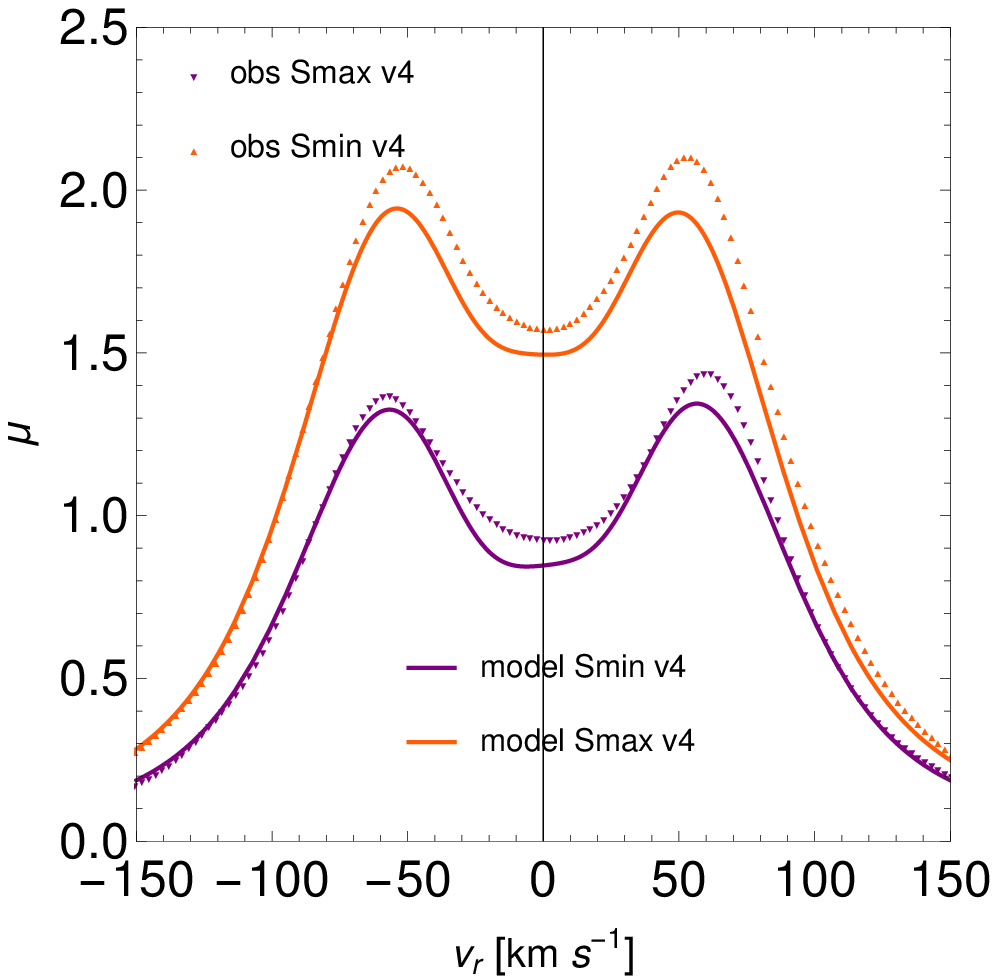}
 
\includegraphics[width=0.49\textwidth]{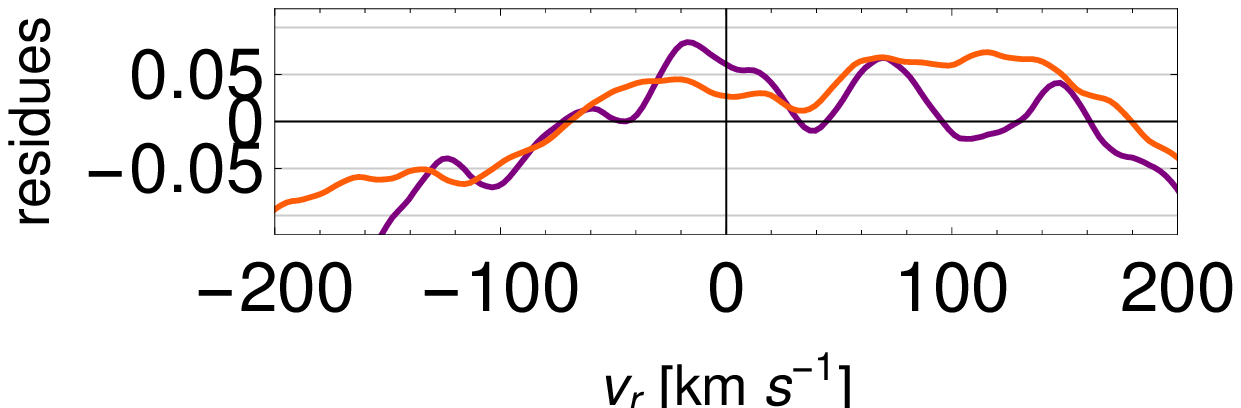}
\includegraphics[width=0.49\textwidth]{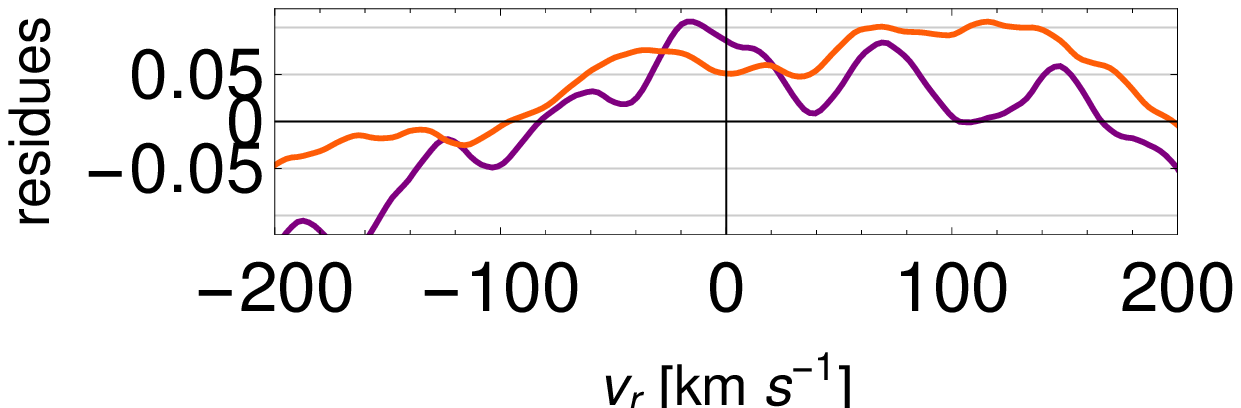}
\caption{Comparison of the profiles based on Version 3 and on Version 4 of the LASP composite \lya{} series. Top left panel: observed profiles taken during solar minimum (Dec. 4, 1996) and solar maximum (Oct. 28, 2001). Gray markers show the original data from \citet{lemaire_etal:15a}, and blue markers represent the re-scaled data. Top right panel: Profiles calculated using our previous model based on Version 3 \iTot{} time series (dashed lines) and those calculated using our new model based on Version 3 \iTot{} time series (solid lines). Middle left panel: Comparison between observed data (points) and our old model (solid lines) for solar minimum (purple) and solar maximum (orange). Middle right panel: Comparison between observed re-scaled data (points) and our new model (solid lines) for solar minimum (purple) and solar maximum (orange). Bottom left panel: Residuals for the model based on Version 3. Bottom right panel: residuals for the model based on Version 4.}
\label{fig:individualFit}
\end{figure*}

The difference between the original and the re-scaled profiles is shown for two example profiles, presented in the top left panel of Figure \ref{fig:individualFit} with grey (Version 3) and blue markers (Version 4). The profiles were taken during solar minimum (Dec. 5, 1996) and solar maximum (Oct. 28, 2001) and they are equivalent to the profiles investigated in our previous paper \citep{IKL:18b}. The original and renormalized profiles taken during the solar maximum in 2001 are almost identical, while the rescaled profiles taken during the solar minimum are systematically different -- Version 4 predicts a stronger line than Version 3. This is easily understandable since the recalibration of the total irradiance (Figure~\ref{fig:Itotv3v4}) resulted in very little modification of \iTot values during solar maximum and a larger change during solar minimum.

\subsection{Updated model parameters} 
\label{sec:model:parameters}
\begin{figure*}
\includegraphics[width=\textwidth]{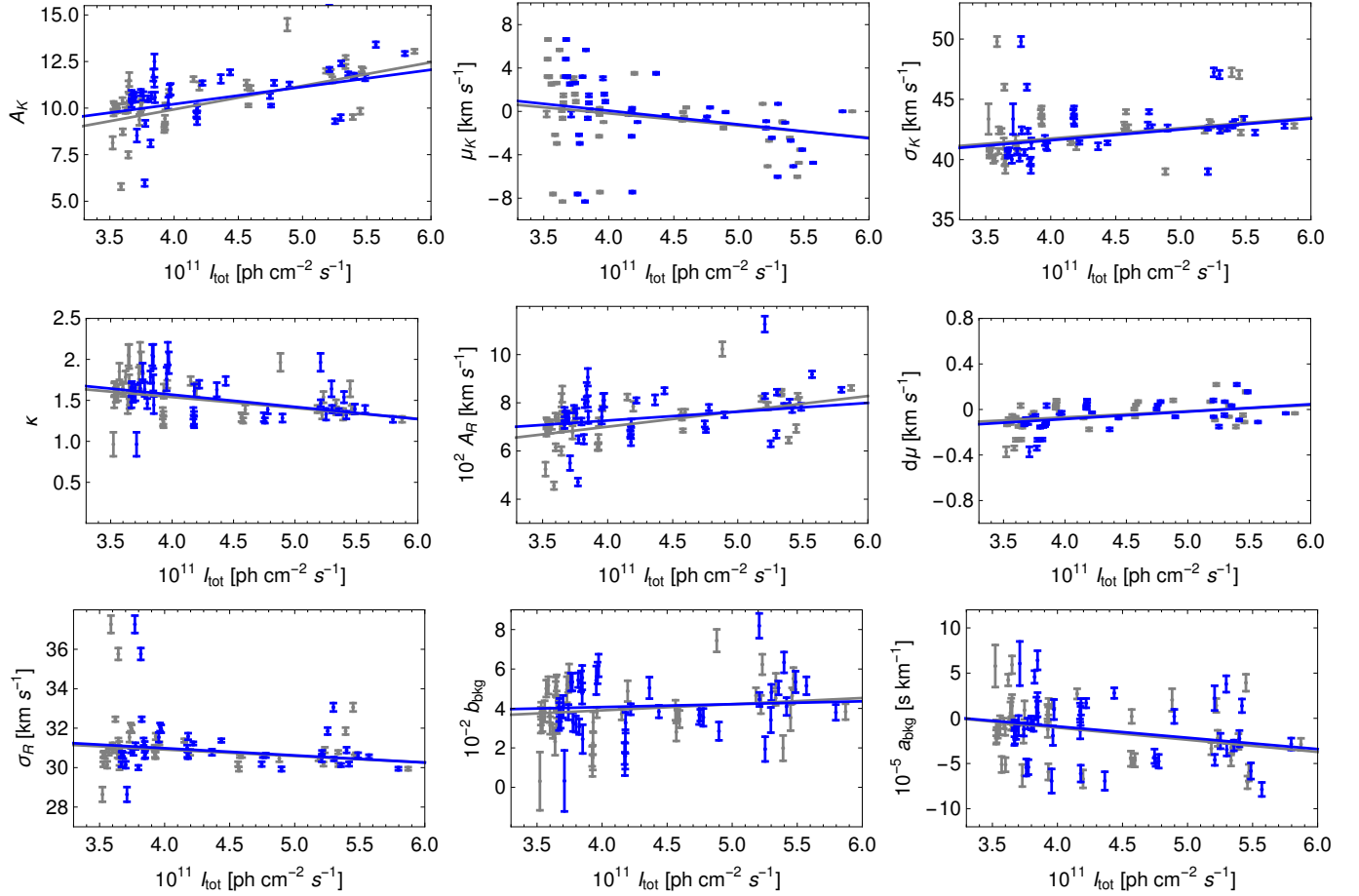}
\caption{Linear correlations between the parameters of the fitted model and \iTot. Each panel shows one of the parameters listed in Table \ref{tab:linCo}. Each point along with its error bars is obtained from a fit to an individual observation of the Lyman-$\alpha$ line profile. Gray points represent the results based on the irradiance calibration Version 3, while the blue ones are obtained using the new Version 4. Lines are fitted to the points using the least squares method.}
\label{fig:linCorr}
\end{figure*}

\noindent
With the original profiles renormalized, we repeated the least-squares fitting of the model parameters defined in Equations 8--11 in \citet{IKL:18a} to all 43 profiles observed by \citet{lemaire_etal:15a}. The best fitting values of the parameters along with the nominal errors of the fitting procedure are shown in Figure ~\ref{fig:linCorr}. Blue points with error bars represent the new values based on Version 4 of the composite \iTot time series, and the gray points, based on Version 3, are shown for comparison. All parameters are plotted as functions of \iTot. Additionally, the linear correlations used to express each parameter as a linear function of \iTot, are shown as solid lines. The numerical values of the coefficients of the linear functions, defined as $P_i=\beta_i \left( 1+ \alpha_i \frac{I_{\mathrm{tot}}}{\langle I_{\mathrm{tot}}\rangle} \right)$, are listed for all parameters in Table~\ref{tab:linCo} for hydrogen and in Table~\ref{tab:linCoD} for deuterium.    

\begin{deluxetable}{ccc}
\tablecaption{\label{tab:linCo}Updated coefficients of the linear correlations between the model parameters for H and the total irradiance in Lyman-$\alpha$, defined in Equation~13 in \citet{IKL:18a}.}
\tablehead{
		\colhead{Parameter ($P_i$)} & \colhead{$\beta_i$} & \colhead{$\alpha_i$}
	}
\startdata
$A_K$ & $  6.523$ & $0.619$ \\
$\mu_K$ & $  5.143$ & $-1.081$ \\
$\sigma_K$ & $ 38.008$ & $0.104$ \\
$\kappa$ & $ 2.165$ & $-0.301$ \\
$A_R$ & $ 580.37$ & $0.28$ \\
$d\mu$ & $ -0.344$ & $-0.828$ \\
$\sigma_R$ & $ 32.439$ & $-0.049$ \\
$b_{bkg}$ & $ 0.035$ & $0.184$ \\
$a_{bkg}$ & $ 0.411\cdot 10^{-4}$ & $-1.333$ \\
\enddata
\tablecomments{ The model along with the parameter values is available online: http://users.cbk.waw.pl/~ikowalska/index.php?content=lya}
\end{deluxetable}
\begin{deluxetable}{ccc}
\tablecaption{\label{tab:linCoD}Updated coefficients of the linear correlations between the model parameters for D and the total irradiance in Lyman-$\alpha$, defined in Equation~13 in \citet{IKL:18a}.}
\tablehead{
		\colhead{Parameter ($P_i$)} & \colhead{$\beta_i$} & \colhead{$\alpha_i$}
	}
\startdata
$A_K$ & $  3.264$ & $0.619$ \\
$\mu_K$ & $-76.237$ & $0.213$ \\
$\sigma_K$ & $ 38.008$ & $0.104$ \\
$\kappa$ & $ 2.165$ & $-0.301$ \\
$A_R$ & $ 290.41$ & $0.28$ \\
$d\mu$ & $ -0.344$ & $-0.828$ \\
$\sigma_R$ & $ 32.439$ & $-0.049$ \\
$b_{bkg}$ & $ 0.017$ & $0.184$ \\
$a_{bkg}$ & $0.206\cdot 10^{-4}$ & $-1.333$ \\
\enddata
\tablecomments{The model along with the parameters is available online: http://users.cbk.waw.pl/~ikowalska/index.php?content=lya}
\end{deluxetable}
The change of \iTot{} affects the strongest the parameters $A_K$, $A_r$, and $b_{\text{bkg}}$, that is the parameters responsible for the general profile shape and the background. $A_K$ is the height of the kappa-component of the profile, $A_r$ is the depth of the central reversal, and $b_{\text{bkg}}$ is the slope of the remnant background in the model. The other parameters changed so little that the modifications of the correlation lines in Figure~\ref{fig:linCorr} are barely visible. The change in $A_K$, $A_R$, and $b_{\text{bkg}}$ is understandable given the results of the \iTot{} update: the contrast between the solar minimum and maximum levels is reduced, so the slope of $A_K(I_{\text{tot}})$ smaller. Similarly, the depth of the central reversal is reduced for larger total intensities, and the spectral background is less sensitive to \iTot. The other parameters of the model, corresponding to the widths of the baseline profile and of the self-reversal, as well as to the spectral shift of the central reversal, are very little affected by the update of line-integrated irradiance.

\section{Effects of the model update on the selected effects inside the heliosphere}
\label{sec:comparison}
\noindent
In this section we briefly compare the effect of updating the radiation pressure model on the density of ISN H in selected locations inside the heliosphere and on the model ISN H flux observed by the Interstellar Boundary Explorer \citep[IBEX; ][]{mccomas_etal:09a}. An extensive study of the sensitivity of various ISN H-related quantities to various aspects of radiation pressure was presented by \citet{IKL:18b}. Here, we show the difference between selected aspects of ISN H inside the heliosphere, simulated using the old and the updated versions of the radiation pressure model. The simulations were done using the nWTPM model of the distribution of ISN H inside the heliosphere. All parameters and other assumptions were identical to those used by \citet{IKL:18b} except for the radiation pressure, which now was based on the model presented in our paper. 
\subsection{ISN H Density}
\begin{figure}
\centering
\includegraphics[width=0.95\columnwidth]{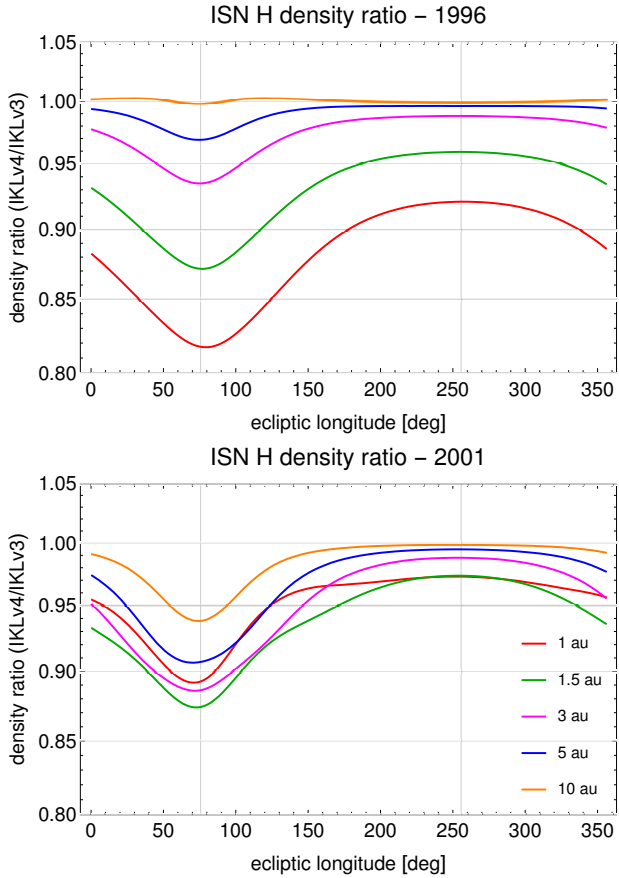}
\caption{Ratio of the hydrogen density based on Version 4 to that based on Version 3 in the ecliptic plane for different distances from the Sun. Top panel shows a minimum of the solar activity in 1996 and bottom panel corresponds to a maximum of solar activity in 2001. The most visible effect is on the Earth's orbit denoted by the red line. Other distances from the Sun are also shown for comparison: 1.5 au in green line, 3 au in magenta line, 5 au in blue line and 10 in orange line. }
\label{fig:figDensity1}
\end{figure}

\begin{figure}
\centering
\includegraphics[width=0.95\columnwidth]{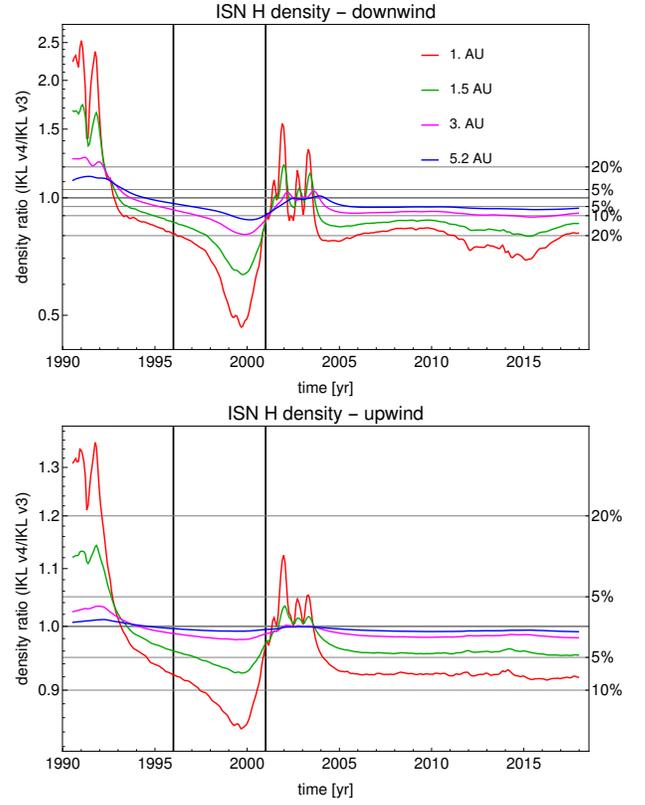}
\caption{Ratio of the hydrogen density based on Version 4 to that based on Version 3 for different distances from the Sun in the downwind (top panel) and upwind directions (bottom panel). Two vertical lines show the solar minimum conditions in 1996 and those for the solar maximum in 2001. The color scheme of the lines is the same as in Figure \ref{fig:figDensity1}.}
\label{fig:figDensityTime}
\end{figure}

Figure~\ref{fig:figDensity1} presents the ratio of the ISN H density based on \iTot{} Version 4 to that based on Version 3 during the minimum (top panel) and maximum of solar activity (bottom panel). The simulations were performed in the ecliptic plane for five distances from the Sun from 1 to 10~au. As it was expected, the biggest effect occurs at 1~au, where in the downwind direction the density based on Version 4 is significantly lower. While in the downwind direction, especially close to the Sun, the hydrogen density is very small, even a slightest change in radiation pressure causes a strong effect on the model density magnitude in this region. Therefore, it is important to use in simulations the most updated and accurate model of radiation pressure, should a precise calculation of the density and related quantities be needed.

In Figure~\ref{fig:figDensityTime}, the ratio of hydrogen density based on Version 4 to that based on Version 3 is shown as a function of time for the downwind (top panel) and upwind (bottom panel) directions. The simulations were performed for the same set of distances from the Sun as for the previous plot. Again, the biggest effect is for the closest distances and in the downwind direction (even up to 50\%). The density change is relatively large inside 2--3~au, where the percentage change is larger than the percentage change in radiation pressure (see bottom panel of Figure~\ref{fig:Itotv3v4}). The sensitivity is larger in the downwind hemisphere. Outside $\sim 3$~au, the effect of the solar \lya{} flux recalibration on the ISN H density becomes negligible.

Results shown in Figure~\ref{fig:figDensityTime} are as expected from analysis of the modification of the magnitude of radiation pressure. When the ratio of \iTot (see bottom panel of Figure~\ref{fig:Itotv3v4}) V4/V3 values is greater than 1, radiation pressure based on Version 4 is stronger, therefore it blows hydrogen away more efficiently and we end up with a lower density (the ratio of densities shown in Figure~\ref{fig:figDensityTime} is less than 1).

\subsection{H$^+$ PUIs}
The change in the ISN H density influence also the H pick up ions (H$^+$ PUI) density. The most affected are PUIs at distances where the ISN H density is the most altered. However, the H$^+$ PUI density is greater than 10\% of the H$^+$ PUI density at the Termination Shock (TS) for distances greater than 1 au \citep{sokol_etal:19b}, where the effect of the change of ISN H density due to the \iTot change is negligible. In consequence, the effect of variation of absolute calibration of \iTot on H$^+$ PUIs is less than 5\% for distances greater than 10 au and thus we can assume it is insignificant, especially at the TS. 

\subsection{Helioglow}
Another quantity potentially affected by the changes of \iTot, and consequently of the radiation pressure is the intensity of the hydrogen backscatter glow. The source function of the backscatter glow is proportional to the magnitude of the solar illuminating flux (\iTot) and the local density of ISN H, and inversely proportional to the square of solar distance. The helioglow intensity is a line of sight integral of the source function. Regions where the source function attains maximum values are located around $\sim 1.5$~au upwind and $\sim 10$~au downwind \citep[see, e.g., Figure~10 in ][]{rucinski_bzowski:95b}. This is largely outside the region strongly affected by the update of the solar flux model. Even though the relative change of the source function close to the Sun may be large (especially in the downwind region because of the large change of the density), its effect on the backscatter glow intensity is expected to be relatively small. A higher \lya{} intensity increases the illumination of ISN H on the one hand, but on the other hand results in an increase of radiation pressure and a decrease of the density. 

\subsection{ISN H flux observed by IBEX-Lo}
\begin{figure*}
\includegraphics[width=\textwidth]{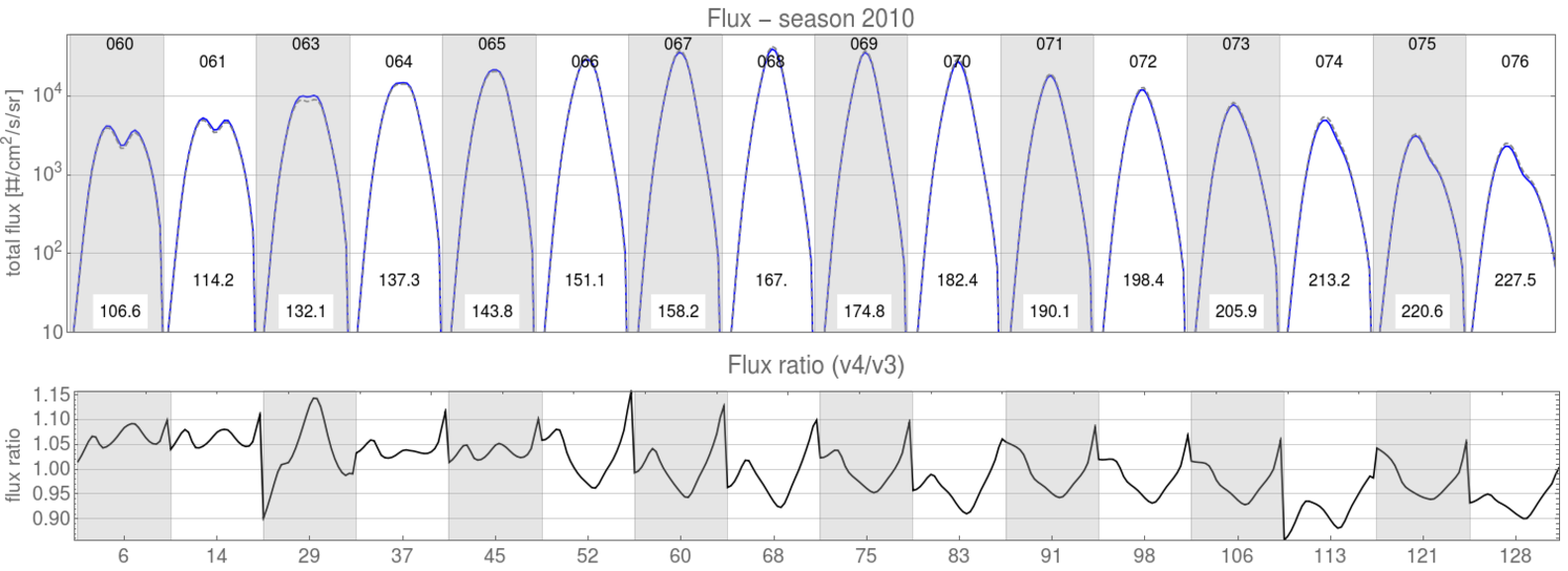}
\includegraphics[width=\textwidth]{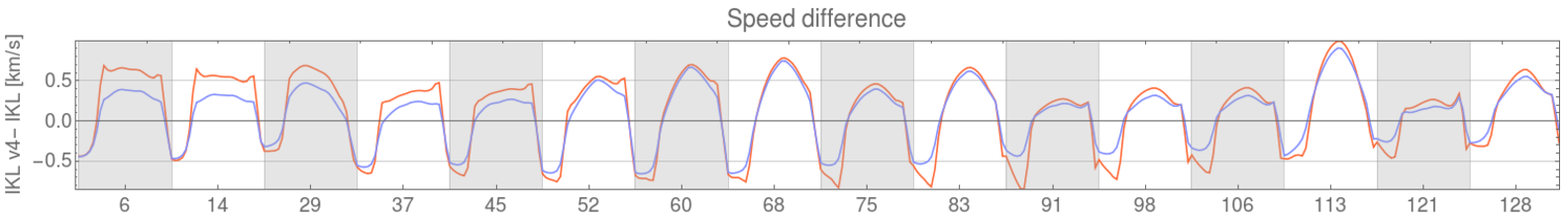}
\includegraphics[width=\textwidth]{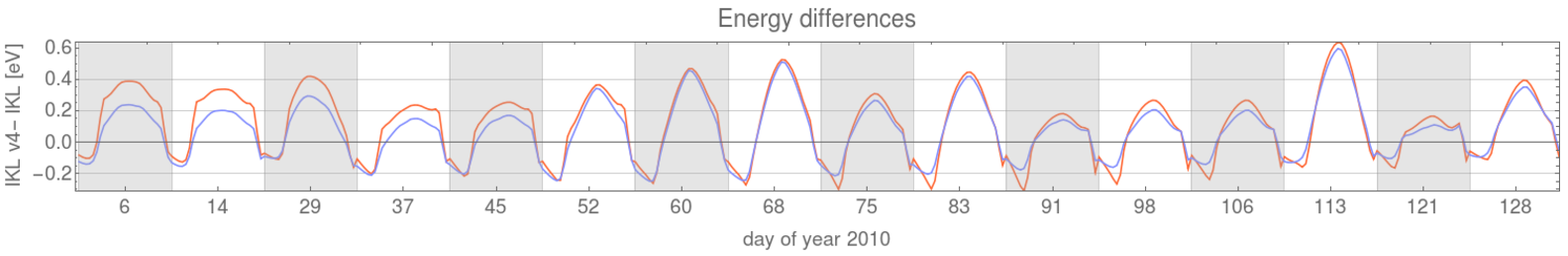}
\caption{Simulated IBEX-Lo signal for ISN H for observation season 2010. First panel shows the total flux (combined primary and secondary populations) based on Version 3 (gray dashed line) and Version 4 (solid blue line). Upper labels show the orbit number, and lower labels correspond to the longitude of the observer. Second panel shows the ratio of the fluxes based on Version 4 and those based on Version 3. The third and fourth panels present a difference between the speed and the energy, respectively, for the two populations separately. Red line is for the primary population, and blue line is for the secondary population). }
\label{fig:IBEX2010}
\end{figure*}

\begin{figure*}
\includegraphics[width=\textwidth]{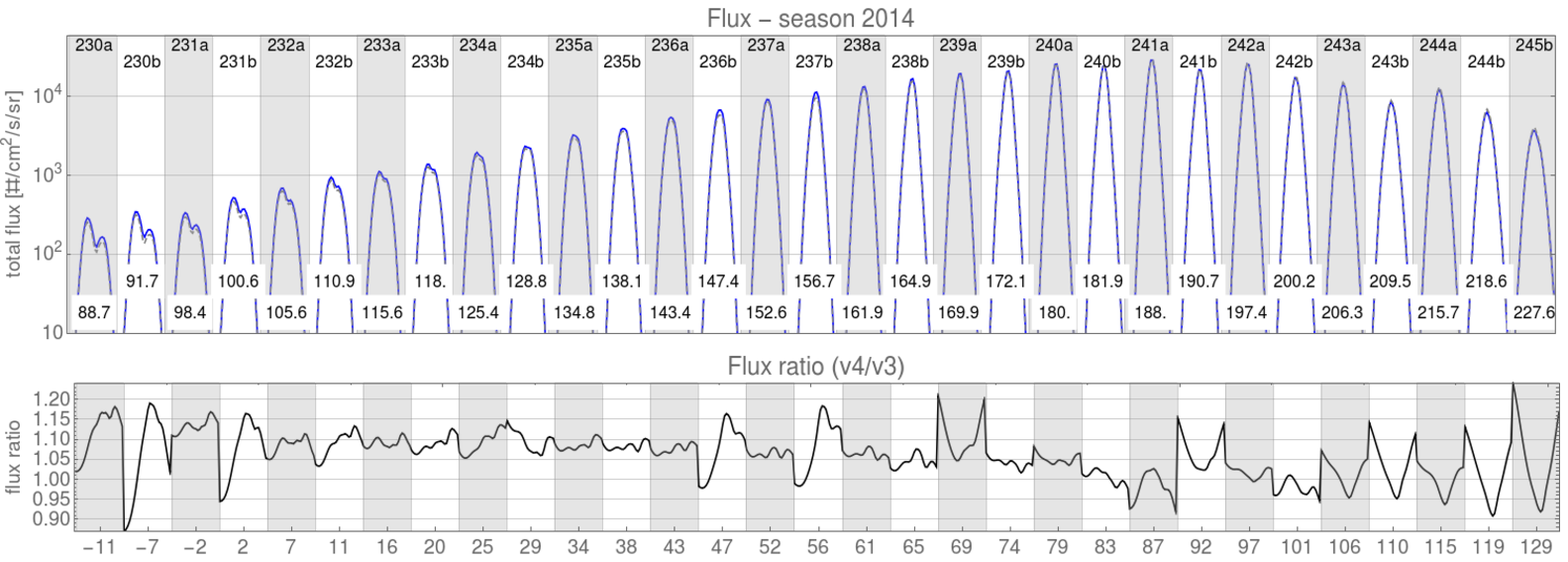}
\includegraphics[width=\textwidth]{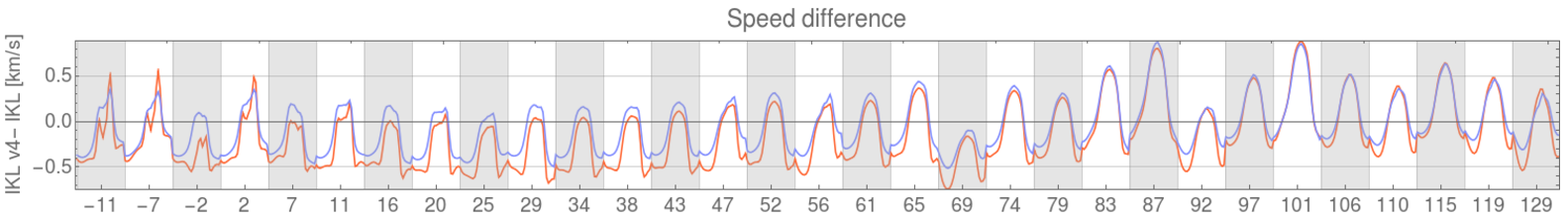}
\includegraphics[width=\textwidth]{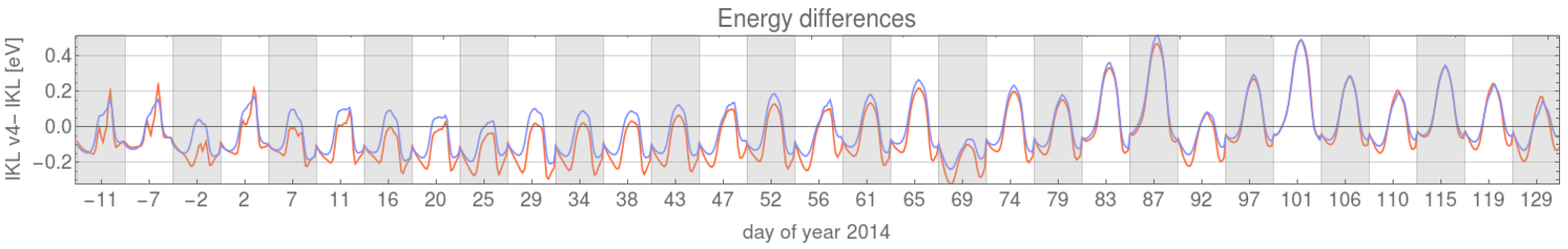}
\caption{Like in Figure~\ref{fig:IBEX2010}, but for observation season 2014.} 
\label{fig:IBEX2014}
\end{figure*}

Yet another aspect where radiation pressure might play an important role is the flux of ISN H at 1~au, which is sampled by IBEX \citep{saul_etal:12a,galli_etal:19a, rahmanifard_etal:19a}. In our previous paper \citep{IKL:18b}, we analyzed the expected differences between the signal simulated using the radiation pressure model by \citet{tarnopolski_bzowski:09} and that by \citet{IKL:18a}. We showed that the effect of this change of radiation pressure model is clearly visible in the simulated signal. Here, we show a similar comparison for the transition from the IKL radiation pressure model based on the solar composite \lya{} flux Version 3 to Version 4.  We made this estimate for the same IBEX-Lo observation seasons as \citet{IKL:18b}: for solar minimum (2010) and solar maximum (2014). The results are shown in Figures \ref{fig:IBEX2010} and \ref{fig:IBEX2014}, respectively. The first panels in these figures present the IBEX-Lo flux based on Version 3 (gray dashed line) and Version 4 (blue solid line). The second panels present the ratio of these quantities. The third and fourth panels show the differences in relative speeds and energies at IBEX-Lo, respectively. The aforementioned quantities are shown for individual IBEX orbits as a function of IBEX spin angle. 

The differences in the flux are largest for the early orbits during the yearly observation seasons, where mostly the secondary ISN H population is observed. However, in this region, there is dominant component of the secondary He population \citep{kubiak_etal:14a}, and the H component has not been clearly identified so far. The magnitude of the ISN H flux differences can be assessed by inspection of the second panel, where a ratio of the fluxes is shown. During solar minimum, the change due to the modification in radiation pressure model varies within (+10\%, -5\%). During solar maximum the change is larger, but the magnitude of the flux has so far precluded its clear detection \citep{saul_etal:13a, galli_etal:19a}. Changes in the relative speed and, consequently, in the relative energy are within 0.5~km~s$^{-1}$ and $\sim 0.4$~eV and are almost negligible.

Throughout the IBEX observation interval (starting at the beginning of 2009), the V4/V3 ratio of \iTot{} is almost constant. Therefore, this study illustrates well the sensitivity of the flux of ISN H to radiation pressure. As shown by \citet{galli_etal:19a}, ISN H is best visible late during the yearly, observation season, when the Earth with IBEX are at ecliptic longitudes 175\degr--200\degr. Within individual orbits, the flux difference varies systematically from $\sim +5$\% to $\sim -5$\% and again back to $\sim +5$\% during solar minimum conditions. This suggests that there is an almost one to one sensitivity of the observed ISN H flux to small variations in radiation pressure. This sensitivity during the solar maximum is of a similar magnitude, even though the behavior of the V4/V3 flux ratios is more complex. 

\subsection{ISN D}
Since the line profile for deuterium is just shifted in radial velocity due to the isotope effect and scaled in the magnitude of radiation pressure due to the mass difference, all above considerations apply to that element as well. The simulated density of deuterium is very small \citep{tarnopolski_bzowski:08a}, and the expected flux at IBEX combined with detection efficiency results in an expected yearly count of detected D atoms at IBEX of just several atoms \citep{kubiak_etal:13a}. Therefore, we will not show detailed analysis of ISN D here. The radiation pressure model parameters for D are listed in Table~\ref{tab:linCoD}.

\section{Summary and conclusions}
\label{sec:summary}
\noindent
Following an update in the absolute calibration of the composite solar \lya{} flux \citep{machol_etal:19a}, we re-evaluated the parameters of the IKL model of solar radiation pressure acting on H and D atoms in the heliosphere \citep{IKL:18a}. The new values of the model coefficients are listed in Table~\ref{tab:linCo} for H and Table~\ref{tab:linCoD} for D.

The updated flux (Figure~\ref{fig:Itotv3v4}) changed by $\pm 10$\%, with occasional spikes to $\pm 20$\%. After $\sim$~2005, the change in \iTot{} is by an almost constant factor of $\sim 4$\%. In the radiation pressure model, the change mostly affects the coefficients responsible for the total height of the profile and for the depth of the central reversal. In general, the contrast between the magnitudes of radiation pressure during the solar maximum and  minimum is slightly reduced. 

We studied the effect of the change in radiation pressure on the distribution of ISN H density inside 10~au from the Sun and on the ISN H flux at 1~au observed by IBEX-Lo. The change in the simulated density may reach as much as 50\% (at 1~au downwind), but is typically much less and fades quickly with increasing solar distance. The IBEX-Lo signal is affected by $\sim 10$\% or less, but in the regions of the Earth orbit where the ISN H signal has been identified, the variation is on the level of $\pm 5$\%. The magnitude of the variation varies from one orbit to another and with the spacecraft spin angle. 

While the changes due to the new calibration of the composite \lya{} flux are mild and only affect regions inside a few au, we recommend adopting the new model of radiation pressure in the heliospheric research, which can be easily implemented and requires only replacing the parameters given by \citet{IKL:18a} with those listed in Table~\ref{tab:linCo} for H and Table~\ref{tab:linCoD} for D.

This analysis can be regarded as a study of the sensitivity to ISN H to variations in radiation pressure. In this respect, a most favorable comparison interval starts in 2005, when the change in radiation pressure is by an almost constant factor of 1.04. We showed that this sensitivity increases with decreasing distance from the Sun  and from the upwind direction towards downwind, as shown in Figure~\ref{fig:figDensityTime}. 

\section*{Acknowledgment}
The authors would like to kindly thank Janet Machol and Martin Snow for providing access to a preprint of their manuscript before it was published. This study was supported by National Science Center, Poland, grants 2018-31-D-ST9-02852 and 2015-19-B-ST9-01328. J.M.S. work was supported by the NAWA Bekker Program Fellowship PPN/BEK/2018/00049.

\bibliographystyle{aasjournal}
\bibliography{iplbib}

\end{document}